\documentclass[a4paper,10pt]{article}
\usepackage[utf8]{inputenc}
\title{Supersymmetric Duality in Deformed Superloop Space }
\author{ Mir Faizal$^1$ and Tsou Sheung Tsun$^2$ \\
$^1$Department of Physics and Astronomy, \\  University of Waterloo,   Waterloo,\\
Ontario N2L 3G1, Canada \\
$^2$Mathematical Institute, University of Oxford,\\ Andrew Wiles
Building, \\Radcliffe Observatory Quarter, Woodstock Road,
\\ Oxford
OX2 6GG, United Kingdom 
 }
 \date{}
\begin{document}
\maketitle
\begin{abstract}
In this paper, we will analyse the superloop space formalism for a 
    four dimensional supersymmetric Yang-Mills theory in deformed superspace. 
We will deform the $\mathcal{N} =1$ superspace by   
imposing non-anticommutativity. This non-anticommutative deformation of the superspace 
will break half the supersymmetry of the original theory. So, this theory will have $\mathcal{N} =1/2$
supersymmetry. We will analyse the 
   superloop space duality  for this deformed 
  supersymmetric Yang-Mills theory using the  $\mathcal{N} =1/2$ superspace formalism.
We will demonstrate that the sources in the original theory will  
become monopoles in the dual theory, and   the monopoles 
in the original theory will become  sources in the dual theory.  
\end{abstract}
\section{Introduction}
It has been observed that certain string theory effects can 
lead to a noncommutative deformation of field theories  \cite{stri}-\cite{stri1}. 
Such noncommutative deformation of ordinary field theories has motivated the study of non-anticommutative 
deformation of supersymmetric field theories   \cite{non}-\cite{non1}. 
The non-anticommutative deformation of  supersymmetric  gauge theories has also been studied
\cite{ng}-\cite{gn}.  
In this deformation, the    Grassmann 
coordinate of a superspace are promoted to   non-anticommutating coordinates. 
Thus, this 
  deformation   breaks the supersymmetry corresponding to those Grassmann variables which are 
  promoted to non-anticommutating coordinates. It is possible to break 
  half the supersymmetry of a four dimensional theory with $\mathcal{N} =1$ supersymmetry. 
 In fact, this deformation has been used for constructing 
 a four dimensional theory with $\mathcal{N} = 1/2$ supersymmetry  \cite{12}-\cite{21}. It is also 
possible to using this deformation to break the supersymmetry of a three dimensional theory. 
As a three dimensional theory with $\mathcal{N} =1$ supersymmetry has only two Grassmann coordinates, 
this deformation will break all the supersymmetry of a three dimensional theory with $\mathcal{N} =1$ supersymmetry. 
However, it is possible to retain some supersymmetry for a  
a three dimensional theory with $\mathcal{N} = 2$. It has been demonstrated that 
a the non-anticommutativity can be used to break the supersymmetry of a three dimensional theory 
 from $\mathcal{N} = 2$ supersymmetry to $\mathcal{N} = 1$ supersymmetry   \cite{az}-\cite{za}.
   In this paper we will analyse the superloop space duality using this  $\mathcal{N} = 1/2$ 
   superspace formalism. This duality is motivated by the loop space duality for ordinary gauge theories. 
   The loop space duality is in turn motivated by the Hodge duality for abelian gauge theories.

There is a duality between the 
electric and magnetic fields that can be constructed using the  Hodge star 
 operation. This duality relies on the fact that the field equation for pure electrodynamics can be  interpreted
as the Bianchi identity for a dual tensor. This dual tensor can be constructed in terms of a dual potential. 
 This duality has been used for analysing various  topological concepts inherent in field theories
  \cite{1a}-\cite{ax11}. 
 This   duality has also been used for analysing 
  many interesting physical phenomena \cite{ax22}-\cite{2a}.
It is also known that the 
  the existence of magnetic monopoles is
equivalent to the quantization of the electric. This is in turn is related to the fact that the
  electromagnetic gauge group is a compact group   \cite{Dirac}. It is not possible to directly 
  generalize this duality to non-abelian gauge theories. This is because the   
  field tensor  for a non-abelian gauge theories is defined as   $F_{\mu\nu} = \partial_\nu A_\mu
- \partial_\mu A_\nu  + ig [A_\mu, A_\nu]$. Now using a  covarient derivative, which is defined as 
$D_\mu = \partial_\mu - ig A_{\mu}$, it is possible to write 
$D^\nu F_{\mu\nu} =0 $, and 
  the Bianchi identity,  ${D^\nu}^\star   F_{\mu\nu} =0 $. However, unlike 
  the abelian case, 
this
does not imply the existence  of a  dual potential. This is  because the covariant derivative 
in the Bianchi identity involves the potential 
$A_\mu$  and not some dual potential $\tilde A_\mu$  corresponding to  $^*   F_{\mu\nu} =0 $. 
However, it is possible to generalize the 
Hodge duality to non-abelian gauge theories using  the  
loop space formalism 
\cite{dual1}.
It has also been demonstrated that this loop space duality reduces to the 
Hodge duality for an  abelian  gauge theory \cite{pq1}-\cite{pq2}.

This loop space duality has been used for analysing aspects of the 
't Hooft's  order-disorder parameters \cite{1t}-\cite{1p}.   
Such a dual potential can be constructed using the loop space formalism. The existence of this  dual potential 
has also motivated the construction of a 
  Dualized Standard Model  \cite{d}-\cite{d0}. The model has been used for analyzing the   off-diagonal
elements of the CKM matrix  \cite{cmk}, and the 
the difference 
  of masses between   different generations of  fermions  \cite{d1}-\cite{d2}. This model has also been 
  used for studding the 
  Neutrino oscillations \cite{no},
and the Lepton transmutations \cite{lt}. 

The loop space formalism is constructed using the 
 Polyakov loops  \cite{p1}. These loops are expressed as   
 the holonomies of closed
loops in space-time. They has also been called as the 
Dirac phase factors in the 
  the physics literature, and they do not depend on the
  parameterization
chosen.  These Polyakov loops are gauge group-valued functions of the
infinite-dimensional loop space. So, no trace is taken over the gauge group. This is what makes the 
  Polyakov loops different from Wilson loops, as in the  Wilson loop a trace is taken  over the gauge group 
  \cite{p1}.  
Thus, unlike the Wilsons loops, the Polyakov loops are elements of the gauge group. 
The   Wilsons loops for super-Yang-Mills theory with
$\mathcal{N} =4$ supersymmetry has been analysed using the superspace formalism \cite{ps}.
Furthermore, the 
 Polyakov loops for three  and four dimensional supersymmetric Yang-Mills theories with 
$\mathcal{N} =1$ supersymmetry  have also been studied \cite{pqaa}-\cite{pq1a}.  
The superloop space duality has also been studied in $\mathcal{N} =1$ superspace \cite{pq1a01}.
In this paper, we will construct such superloops for deformed superspace. Then we will analyse the superloop 
space duality for the gauge theories using this formalism.

\section{Deformed Superloop Space}
In this section, we will analyse   a four dimensional   gauge theory in $\mathcal{N} = 1/2$  
 superspace formalism. Let us start by defining the coordinates of the superspace as 
 $(x^\mu, \theta^a, \bar \theta^{\dot{a}})$. Here $\mu = 0, 1,2, 3$, and $a, \dot{a} = 1, 2$. 
The non-anticommutative deformation can be performed  
by promoting the Grassmann coordinate $\theta^a$ to a non-anticommutating variables, such that 
\begin{equation}
 \{\theta^a, \theta^{b} \} = C^{ab}. 
\end{equation}
Here the product of superfields of $\theta^a$ is Weyl ordered by replacing the ordinary product 
of superfields on the deformed superspace by a star product. 
This star product is the fermionic version of the Moyal product. 
Thus, for two  supervector fields $V(x, \theta, \bar \theta)$, and $V' (x, \theta, \bar \theta)$, 
we have 
\begin{equation}
V(x, \theta, \bar \theta) \star   V'(x, \theta, \bar \theta)= V(x, \theta, \bar \theta) 
\exp \left(-\frac{ C^{ab}}{2} \overleftarrow{\frac{\partial}{\partial \theta^a}} 
\overrightarrow{\frac{\partial}{\partial \theta^b}} \right) V'(x, \theta, \bar \theta).
\end{equation}
The Grassmann coordinate $\bar \theta_{\dot{a}}$ satisfy the usual  relations, 
\begin{eqnarray}
 \{ \bar \theta_{\dot{a}}, \bar \theta_{\dot{b}}\} =0, &&  
 \{ \bar \theta_{\dot{a}},  \theta_{a}\} =0, \nonumber \\
 { [\bar \theta_{\dot{a}}, x^\mu] }=0. &&
\end{eqnarray}
However, we have 
\begin{eqnarray}
 [x^\mu, x^\nu] = \bar \theta \bar \theta C^{\mu \nu} , && [x^\mu, \theta^a] 
 = i C^{ab}\sigma^\mu_{b\dot{b}}\bar\theta^{\dot{b}}, 
\end{eqnarray}
where $ C^{\mu\nu} = C^{ab}  \epsilon_{bd}(\sigma^{\mu\nu})_a^d $. 
Now   we can define  
$
  y^\mu = x^\mu + i \theta^a \sigma^\mu_{a\dot{a}} \bar \theta^{\dot{a}},
$ and  obtain 
\begin{eqnarray}
[  \theta_{a }, y^\mu] =0, &&  
[ \bar \theta_{\dot{a}}, y^\mu] =0 , \nonumber \\
  {[y^\mu, y^\nu]} =0.  &&
\end{eqnarray}
Thus, we can take the superfields to be functions of $(y^\mu, \theta^a, \bar \theta^{\dot{a}})$ 
\cite{non}-\cite{non1}.
Now we can write a supervector field $V(y, \theta, \bar \theta)$ in the Wess-Zumino gauge as 
\begin{eqnarray}
 V(y, \theta, \bar \theta) &=& - \theta \sigma^\mu \bar \theta A_\mu  + i \theta\theta \bar \theta \bar \lambda
 - i    \bar \theta \bar \theta \theta^a \left( \lambda_a + \frac{1}{4}\epsilon_{ab} C^{bd} \sigma^\mu_{d\dot{d}}
 [\bar\lambda^{\dot{d}}, A_\mu] \right) \nonumber \\ && + \frac{1}{2} 
 \theta\theta \bar \theta \bar \theta  (D - i \partial_\mu A^\mu).  
\end{eqnarray}
Here we have defined $V^A(y, \theta, \bar \theta)T_A = V(y, \theta, \bar \theta)$,
with    $[T_A, T_B] = i f_{AB}^CT_C$.
The Chiral and anti-Chiral field strength for the gauge theory are defined to be 
$4 W_a = - \bar D  \bar D e^{-V}_\star  \star D_a e^V_\star$ 
and $4 \bar W_{\dot{a}} =   D    D e^{-V}_\star  \star \bar D_{\dot{a}} e^V_\star$,  respectively. Now the
   Lagrangian  for the deformed gauge theory can be written as \cite{ng}-\cite{gn}
\begin{equation}
 \mathcal{L} = Tr \int d^2 \theta  \, W^a \star   W_a + Tr \int d^2\bar \theta \,  \bar W^{\dot{a}} \star   \bar W_{\dot{a}}.   
\end{equation}
In component form this can be written as 
\begin{eqnarray}
  \mathcal{L} &=& Tr ( - 4 i \bar \lambda \sigma^\mu D_\mu \lambda - F^{\mu\nu} F_{\mu\nu} + 2 D^2) 
 \nonumber \\ &&  + Tr\left( - 2 iC^{\mu\nu} F_{\mu\nu}\bar\lambda\bar \lambda  +
  \frac{C^{\mu\nu}C_{\mu\nu}}{2} (\bar \lambda\bar \lambda)^2 \right). 
\end{eqnarray}
It is also possible to express this  deformed four dimensional supergauge theory using 
  covariant derivative  defined as  \cite{1001} 
  \begin{eqnarray}
 \nabla_A &=& (-i \{\mathcal{D}_a , D_{\dot{a}}\}_\star  , 
\mathcal{D}_a , D_{\dot{a}}), 
 \nonumber \\ 
 \exp ( V )_\star \star \nabla_A \star \exp( -V )_\star &=& (-i \{ D_a, \mathcal{D}_{\dot{a}} \}_\star  , 
  D_a , \mathcal{D}_{\dot{a}} ), 
\end{eqnarray}
where $\mathcal{D}_a = \exp (-V)_\star \star   D_a \exp (V )_\star$ and $\mathcal{D}_{\dot{a}} =
\exp (V)_\star \star   D_{\dot{a}} \exp(- V )_\star$.  It is also possible to express this covarient derivative as
$ \nabla_A = D_A - i \Gamma_A$ \cite{pq1a}.
Here the superspace derivative $D_A$ is defined by 
$D_A = ( \partial_{a\dot{a}}, D_a, D_{\dot{a}})$ and the superspace connection $\Gamma_A$ is defined by 
and $\Gamma_A = (\Gamma_{a\dot{a}},\Gamma_a, \Gamma_{\dot{a}} )$. 
We can define $H_{AB}=  [ \nabla_A, \nabla_B \}_\star   = T^C_{AB}\nabla_C - i F_{AB} $, and then the 
  Bianchi identity  will be written as 
$ [\nabla_{[A}, H_{BC)}\}_\star    =0.$ 

The covariant derivative  transforms under gauge transformation as 
$\nabla_A \to e^{ i \Lambda}_\star \star \nabla_A \star e^{-i\Lambda}_\star$, 
and $ e^{V}_\star \star \nabla_A \star e^{-V}_\star  
\to e^{i \bar \Lambda}_\star \star e^{V}_\star \star  \nabla_A \star e^{-V}_\star \star e^{ -i \bar\Lambda}_\star$. 
It is possible to use another representation in which the   
  covariant derivatives    transform under gauge transformations  
as $\nabla_A \to u \star \nabla_A \star u^{-1}$ \cite{1001}. Here we have defined $u$ as 
  $ u = e^{iK}_\star $ where  parameter $K = K^A T_A$ is a real superfield. 
Now the transformation of the  spinor fields can be expressed as  $\Gamma_a \to i u \star \nabla_a \star u^{-1}, 
  \Gamma_{\dot{a}} \to i u\star    \nabla_{\dot{a}}\star u^{-1}$ and 
 $ \Gamma_{a\dot{a}} \to i u \star    \nabla_{a \dot{a}} \star u^{-1}$.

Now we can derive the duality for this deformed superspace. This can be done by first using the conventional 
constraints as  
$F_{a\dot{a}} = F_{ab} = F_{\dot{a}\dot{b}} =0$.
As the super-connection is defined by $\Gamma_A = (\Gamma_{a\dot{a}},\Gamma_a, \Gamma_{\dot{a}} )$, we 
can parameterize the superloop by 
 $ \xi (s)= (\sigma^\mu \xi_\mu(s))^{a\dot{a}}\theta_{a}\theta_{\dot{a}} + 
\xi^a (s)\theta_a + \xi^{\dot{a}} (s)\theta_{\dot{a}} $, and so we can write 
$\xi^A = (\xi^{a\dot{a}}, \xi^a, \xi^{\dot{a}})$ \cite{1001}.
It may be noted that for higher dimensional theories, and for theories with higher amount of supersymmetry 
we will have to choose a different parameterization.  Now we can 
 parameterized the superloop     along a curve $C$ as 
\begin{equation}
 C : \{ \xi^A (s): s = 0 \to 2\pi, \, \, \xi^A (0) = \xi^A(2\pi)\},  
\end{equation}
 where $\xi^A (0) = \xi^A(2\pi)$ is a fixed point on this curve \cite{pq1a}. We can now define the superloop 
 variable for the deformed superspace as
\begin{eqnarray}
 \Phi [\xi] &=& 
 P_s \exp i \int^{2\pi}_0 \left[ \Gamma^{a\dot{a}} (\xi(s))  \frac{d \xi_{a\dot{a}}(s)}{ds} +  \Gamma^a (\xi(s))  
\frac{d \xi_a(s)}{ds} \right. \nonumber \\  && \left. +   \Gamma^{\dot{a}} (\xi(s))  
\frac{d \xi_{\dot{a}}(s)}{ds}\right]_\star   
 \nonumber \\  &=&
 P_s  \exp i \int^{2\pi}_0  \left[\Gamma^A (\xi(s)) \frac{d \xi_A(s)}{ds} \right]_\star  . 
\end{eqnarray}
where all the products are taken as star products. Furthermore,   $P_s$ denotes ordering in $s$. Here 
this ordering is 
increasing from right to left. The  derivative in $s$ is  taken from below.
This superloop space is a scale superfield on the deformed superspace from 
  the supersymmetric point of 
view. Thus, it has $\mathcal{N} = 1/2$ supersymmetry. 

The parallel transport between two points, $\xi(s_1)$ and $\xi(s_2)$, 
can be defined as 
\begin{eqnarray}
 \Phi [\xi: s_1, s_2 ] &=& 
 P_s \exp i \int^{s_2}_{s_1} \left[ \Gamma^{a\dot{a}} (\xi(s))  \frac{d \xi_{a\dot{a}}(s)}{ds} +  \Gamma^a (\xi(s))  
\frac{d \xi_a(s)}{ds} \right. \nonumber \\  && \left. +   \Gamma^{\dot{a}} (\xi(s))  
\frac{d \xi_{\dot{a}}(s)}{ds}\right]_\star   
 \nonumber \\  &=&
 P_s  \exp i \int^{s_2}_{s_1}  \left[\Gamma^A (\xi(s)) \frac{d \xi_A(s)}{ds} \right]_\star  . 
\end{eqnarray}
Here it is defined along path parametrized by $\xi$. 
It is possible to use $\Phi[\xi]$ to define  $F_A[\xi|s]$  
\begin{eqnarray}
 F_A [\xi| s] &=& 
 i \Phi^{-1}[\xi]\star   \delta_A (s) \star   \Phi[\xi]
  \nonumber \\
 &=& \Phi^{-1}[\xi: s,0] \star   H^{AB} (\xi (s) ) \star   \Phi  [\xi: s,0] \star   \frac{d \xi_B (s) }{d s},  
\end{eqnarray}
where $\delta_A (s) = \delta /\delta \xi^A (s)= 
(\delta /\delta \xi^{a\dot{a}}(s),\delta /\delta \xi^a (s), \delta /\delta \xi^{\dot{a}} (s)))  $.
Here we start from a point and along a  path  till we reach the point 
$s$, and then we return to the original point along the same point. 
The phase factor   from the original point to  $s$ cancels the 
phase factor a  from $s$ to the original point. However, we do obtain an additional contribution for 
 $H^{AB}(\xi (s) )$   due to the  infinitesimal circuit 
generated at $s$.

\section{Deformed Superloop Space Duality }
It is possible to write the duality using   loop space formalism for ordinary non-abelian 
gauge theories \cite{pq1}-\cite{pq2}.
Here we will generalize this duality to deformed superspace. In order to that we will first 
analyse  function curl and divergence of a superloop variable. 
We can define a  functional curl and a functional divergence as 
\begin{eqnarray}
 (\rm{curl}\,  F [\xi|s])_{AB} &=& \delta _A (s)   F_B[\xi|s] -
\delta _B (s)   F_A[\xi|s], \nonumber \\
\rm{div}\,  F [\xi|s] &=& \delta ^A (s)  F_A[\xi|s].
\end{eqnarray}
As the superloop variables are highly redundant, we need to  constrained them by an infinite set of conditions. 
These can be  expressed by the vanishing of the superloop space curvature  \cite{pq1a},
$G_{AB}[\xi, s] =  (\rm{curl}\,  F [\xi|s])_{AB}
+i [F_A [\xi|s], F_B [\xi|s]]_\star  
= 0 $. 
We can also define $-iG_{AB}[\xi, s]$ as a commutator of two covarient  superloop derivatives, 
$[\nabla_A [\xi, s], \nabla_B [\xi, s]]_\star$, where 
$
 \nabla_A [\xi, s] =  \delta_A (s)  -i F_A [\xi| s]. 
$

It is possible to construct  $E_A [\xi|s]$ from $F_A [\xi|s]$, 
\begin{equation}
 E_A [\xi|s] =  \Phi[ \xi: s, 0] \star   F_A[\xi|s] \star    \Phi^{-1}[ \xi: s, 0], 
\end{equation}
Thus, we can construct $E_A [\xi|s]$ from  $F_A [\xi|s]$ using parallel transport.  
Now as the 
  $E_A [\xi|s]$ only depends  on a segment of the loop  $\xi(s)$ around $s$, it is  a 
segmental variable rather than a full superloop variable.   
Now as the integrals involving $E_C [\xi|s]$ depends   will depend on the a little segment from $s_-$ to $s_+$, so   
    limit $\epsilon \to 0$ can only be  taken only after integration. 
    Here we have defined $\epsilon = s_+ - s_- $. 
As 
 segment can shrinks to a point, and   we can write $E^A [\xi|s] \to H^{AB} (\xi (s)) \star   d \xi_B (s) /d s$. 
In fact,  all the loop operations require a segment of the loop on which they can operate, 
so this limit can only be taken  after all the superloop operations have  been
performed. Now we can define  
\begin{eqnarray}
 (\rm{curl}\, E [\xi|s])_{AB} &=& \delta _A (s) E_B[\xi|s] -
\delta _B (s) E_A[\xi|s], \nonumber \\
\rm{div}\, E [\xi|s] &=& \delta ^A (s) E_A[\xi|s].
\end{eqnarray}
Thus, we obtain 
\begin{eqnarray}
 \delta_A (s') E_B[\xi|s] &=&  \Phi[ \xi: s, 0] \star   [ \delta_A (s') F_B[\xi|s] 
 \nonumber \\ && + i \Theta (s-s') [F_A [\xi|s], F_B [\xi|s]]_\star   ] \star   \Phi^{-1}[ \xi: s, 0], 
\end{eqnarray}
where $i \Theta (s-s')$ is the Heavisde function. 
So,  the superloop space curvature can now be written as $ G_{AB }[\xi, s] = 
\Phi[ \xi: s, 0] \star   ({\rm{curl}} E [\xi|s] )_{AB} \star   \Phi^{-1}[ \xi: s, 0]$ and thus the 
constraints can be fixed as $({\rm{curl}}E [\xi|s] )_{AB} =0$. 

Now we can define a new 
  variable 
$\tilde E_A [ \eta|t]$ which is dual to 
$E_A [\xi|s]$. Now using  $\eta (t)$ as another parameter superloop, we can write 
\begin{eqnarray}
 \omega^{-1}   [\eta(t) ] \star  \tilde E^A [ \eta|t] \star   \omega [\eta(t)]&=&  
 - \frac{2}{N} \epsilon^{ABCD}\frac{d\eta_B(t)}{dt} \star   \int  D\xi ds \delta(\xi (s) - \eta(t)) \nonumber \\  
 && \star    E_C [\xi|s] \star   \frac{d\xi_D (s)}{ds} 
 \nonumber \\ && \star   \left[\frac{d\xi^F (s)}{ds} \star   \frac{d\xi_F(s) }{ds} \right]^{-2} , \label{1d}
\end{eqnarray}
where $N$ is a normalization constant.  
Here the  local rotational matrix is denoted by  $ \omega [\eta(t) ] $. This corresponded to 
  transforming the quantities from a direct frame to the dual frame.  
 The   loop derivative of $\tilde E^A [ \eta|t]$ can be calculated by 
 using the fact that $\tilde E^A [ \eta|t]$ is a segmental quantity. It depends on a segment from $t_-$
to $t_+$. Here again we take the    limit $\epsilon' \to 0$ only  after all the superloop operations have  been
performed. Here we have defined $\epsilon' = t+- t_-$.  
We also have  $\epsilon' < \epsilon$. The 
  $\delta$-function    ensures the variable $\xi (s)$ coincides with $\eta(t)$ from 
$s = t_-$ to $s = t_+$. As
the segment shrinks to a point, 
and we obtain $E^A [\eta|t] \to \tilde H^{AB} (\eta(t)) d \eta_B (t) /d t$. 

It may be noted that we can define the gauge transformation of  $E_A[\xi|s]$
and $\tilde E^A [ \eta|t]$ as
\begin{eqnarray}
  E_A[\xi|s] &=& [1 + i \Lambda [\xi(s)] ]  \star  E_A[\xi|s]\star [1 - i \Lambda [\xi(s)] ], 
\nonumber \\
\tilde E^A [ \eta|t]&=& [1 + i \tilde \Lambda [\eta(t)] ] \star\tilde   E_A[\eta|t]\star[1 - i \tilde \Lambda [\eta(t)] ].  
\end{eqnarray}
 Here the gauge parameters $\Lambda [\xi(s)]$ and $\tilde \Lambda [\eta(t)]$ have zero loop derivatives. 
Here  the  dual quantity     
  $\tilde H^{AB}$ can be constructed from a dual 
potential, where  $\tilde \Gamma_A = (\tilde \Gamma_{a\dot{a}},\tilde\Gamma_a, \tilde 
\Gamma_{\dot{a}})$,  and  $ [\tilde \nabla_A, \tilde\nabla_B \}_\star   =  \tilde H_{AB} $. Here we have defined 
 $ \tilde\nabla_A = D_A - i \tilde \Gamma_A$. 
 The dual covariant derivative  transforms under gauge transformation as  
  $\tilde \nabla_A \to \tilde u \star \tilde \nabla_A \star \tilde u^{-1}$. Here we have defined $\tilde u$ as 
  $ \tilde u = e^{i\tilde K}_\star $ where  parameter $\tilde K  $ is a real superfield. 
Now the transformation of the  spinor fields can be expressed as  $\tilde \Gamma_a \to i \tilde u \star 
\tilde \nabla_a \star \tilde u^{-1}, 
 \tilde \Gamma_{\dot{a}} \to i \tilde u\star   \tilde  \nabla_{\dot{a}}\star \tilde u^{-1}$ and 
 $\tilde \Gamma_{a\dot{a}} \to i \tilde u \star  \tilde   \nabla_{a \dot{a}} \star \tilde u^{-1}$.
 It is also possible to use a different representation under which the dual covarient derivative transform as 
$\tilde \nabla_A \to  e^{ i \tilde \Lambda}_\star \star \tilde \nabla_A \star e^{-i\tilde \Lambda}_\star$, 
and $ e^{\tilde V}_\star  \star \tilde \nabla_A \star e^{-\tilde V}_\star  
\to e^{i  \tilde {\bar\Lambda}}_\star \star e^{\tilde V}_\star \star  \tilde \nabla_A \star e^{-\tilde V}_\star
\star e^{ -i \tilde{\bar\Lambda}}_\star$. 
 We can again define the dual covarient derivative in terms of a dual supervector field $\tilde V$
 \begin{eqnarray}
 \tilde \nabla_A &=& (-i \{\tilde \mathcal{D}_a ,   D_{\dot{a}}\}_\star  , 
\tilde \mathcal{D}_a ,  D_{\dot{a}}), 
 \nonumber \\ 
 \exp ( \tilde V )_\star \star \tilde \nabla_A \star   \exp( -\tilde V )_\star &=& (-i \{  D_a, 
 \tilde \mathcal{D}_{\dot{a}} \}_\star  , 
 D_a , \tilde \mathcal{D}_{\dot{a}} ), 
\end{eqnarray}
where $\tilde \mathcal{D}_a = \exp (-\tilde V)_\star \star   D_a \exp (\tilde V )_\star$ 
and $\tilde \mathcal{D}_{\dot{a}} =
\exp (\tilde V)_\star \star   D_{\dot{a}} \exp(- \tilde V )_\star$. 
Now the dual supervector field $ \tilde V(y, \theta, \bar \theta)$ can be written as 
 \begin{eqnarray}
\tilde  V(y, \theta, \bar \theta) &=& - \theta \sigma^\mu \bar \theta \tilde A_\mu  
 + i \theta\theta \bar \theta \tilde{\bar \lambda}
 - i    \bar \theta \bar \theta \theta^a \left( \tilde \lambda_a + \frac{1}{4}\epsilon_{ab} C^{bd} 
 \sigma^\mu_{d\dot{d}}
 [\tilde{\bar\lambda}^{\dot{d}} , \tilde A_\mu] \right) \nonumber \\ && + \frac{1}{2} 
 \theta\theta \bar \theta \bar \theta  (\tilde D - i \partial_\mu \tilde A^\mu).  
\end{eqnarray}
It may be noted that the dual supervector potential is also a function of deformed superspace coordinates. 
Thus, for two dual supervector fields $\tilde V(y, \theta, \bar \theta)$, and $\tilde V' (y, \theta, \bar \theta)$, 
we have 
\begin{equation}
\tilde V(y, \theta, \bar \theta) \star  \tilde V'(y, \theta, \bar \theta)= \tilde V(y, \theta, \bar \theta) 
\exp \left(-\frac{ C^{ab}}{2} \overleftarrow{\frac{\partial}{\partial \theta^a}} 
\overrightarrow{\frac{\partial}{\partial \theta^b}} \right) \tilde V'(y, \theta, \bar \theta).
\end{equation}
It is also possible to define the 
Chiral and anti-Chiral field strength for the dual theory as 
$4 \tilde W_a = - \bar D  \bar D e^{-\tilde V}_\star \star  D_a e^{ \tilde V}_\star$ 
and $4 \tilde{\bar W}_{\dot{a}} =   D    D e^{-\tilde V}_\star \star \bar D_{\dot{a}} e^{\tilde V}_\star$,  
respectively.

\section{Application of Duality}
In this section, we will demonstrate that the sources of the ordinary theory becomes monopoles in the dual theory, 
and the monopoles in the dual theory become sources in the ordinary theory.  Before doing that we note that this 
duality reduces to   an ordinary superloop space duality if we neglect the effect of 
noncommutativity \cite{pq1a01}. Furthermore, for if for the non-supersymmetric case, this reduces to the ordinary 
loop space duality. Thus, if we use 
  $[\Phi[\xi]]_| = \phi[\xi]$ as the loop space variable, then we can obtain  
  $E_\mu [\xi|s]$ from $F_\mu [\xi|s]$, where 
$F_\mu [\xi|s]$ is the loop space connection  corresponding to loop variable $[\Phi[\xi]]_| = \phi[\xi]$.
Now in absence of non-anticommutative deformation, we can construct the dual variable to 
the usual loop space variable $E_\tau [\xi|s]$ as  $\tilde E^\mu [ \eta|t] $. 
Then in the limit 
in which the     width of $\tilde E^\mu [ \eta|t]$ going to zero, we can  show that  
$
 \tilde F^{\mu\nu} [x]  = 
 - \epsilon^{\mu\nu\tau \rho} F_{\tau \rho} [x]/2 
$ \cite{pq2}. 
Thus, the usual  Hodge star operation can be obtained by identifying $\tilde F_{\mu\nu}$ with $^*   F_{\mu\nu}$. 
It may be noted that   the usual  loop space variable can be used for analysing the 
't Hooft's  order-disorder parameters \cite{p1}.   These order-disorder parameters 
can be constructed by using 
  two spatial loops $C$ and $C'$ with the linking number $n$ between them. Here $su(N)$ is the 
which is used for this analysis. The magnetic flux through $C$ is
measured by $A(C)$, and  the electric flux through $C$ is measured by 
  $B(C)$. The 
  order-disorder 
parameters are defined as  $A(C) B(C') = B(C') A(C)  \exp (2\pi in/N)$. 
Now   $A(C)$ is expressed using the    potential $A_\mu$ and 
$B(C)$ is expressed using the dual potential  $\tilde A_\mu$
\cite{d}-\cite{1p}.

Now we will demonstrate that the duality transformation is invertible. This can be done by writing a duality 
transformation for 
   $E^A [ \zeta|u]$ as, 
\begin{eqnarray}
 &&\omega^{-1} \star   [\zeta(u) ]E^A [ \zeta|u] \star   \omega [\zeta(u)]\nonumber \\ &=&  
 - \frac{2}{N} \epsilon^{ABCD}\frac{d\zeta_B(u)}{du} \star   \int  D\eta dt \tilde E_C [\eta|t] \star  
 \frac{d\eta_D (t)}{dt} \nonumber \\ &&  
 \star   \left[\frac{d\eta^F (t) }{dt} \star   \frac{d\eta_F (t) }{dt} \right]^{-2} \delta(\eta (t) - \zeta(u)), 
\end{eqnarray}
where $\zeta_B (u)$ is a new loop parameterized by $u$.
So, we can write  $A^A [\zeta (u) ]$ as 
\begin{eqnarray} A^A [\zeta (u) ]&=&
 \frac{2}{N} \epsilon^{ABCD}\frac{d\zeta_B(u)}{du} \star   \int  D\eta dt \omega^{-1}   [\eta(t) ] \star   \tilde E_C [\eta|t] 
 \star   \omega [\eta(t) ] \nonumber \\ && \star  
 \frac{d\eta_D (t)}{dt} \star  
 \left[\frac{d\eta^F (t) }{dt}\star   \frac{d\eta_F (t) }{dt} \right]^{-2} \delta(\eta (t) - \zeta(u))
 \nonumber \\ & =& 
 -\frac{4}{N} 
  \epsilon^{ABCD}\frac{d\zeta_B(u)}{du} \star   \int  D\eta D \xi  dt ds  
 \frac{d\eta_D(t)}{dt} \star   \frac{d\eta^Q(t)}{dt} \nonumber \\ &&  
 \star   \left[\frac{d\eta^X (t) }{dt} \star   \frac{d\eta_X (t) }{dt} \right]^{-2}
 \delta(\eta(t) - \zeta (u)) \star   E^W [\xi|s]  \nonumber \\ && 
\star   \frac{d\xi^E(s)}{ds} \star   
 \left[\frac{d\xi^Y (s) }{ds}\star   \frac{d\xi_Y (s) }{ds} \right]^{-2} \delta(\xi (s) - \eta(t))
 \epsilon_{CQWE}.
\end{eqnarray}
Now we can write  
\begin{eqnarray}
 && \omega^{-1} [\zeta(u) ] \star   E^A [ \zeta|u]\star    \omega [\zeta(u)]\nonumber \\ &=&  
 - \frac{2}{N} \epsilon^{ABCD}\frac{d\zeta_B(u)}{du}   \star   \int  D\eta dt \tilde E_C [\eta|t] \star  
 \frac{d\eta_D (t)}{dt} \nonumber \\ &&  
\star   \left[\frac{d\eta^F (t)}{dt}\star   \frac{d\eta_F(t) }{dt} \right]^{-2} \delta(\eta (t) - \zeta(u)). \label{2d}
\end{eqnarray}
Thus, by  identifying $\zeta (u) $ with $\xi(s)$, we obtain the desired result that  
 this duality is invertible.  
 
The     source term in the deformed supersymmetric Yang-Mills theory can be defined as
 $\nabla^C \star   H_{BC} \neq 0 $ and
  $ {\rm{div}}F [\xi |s] \neq 0$.  As we have 
${\rm{div}}E [\xi |s] = \Phi[ \xi: s_1, 0] \star   {\rm{div}}F [\xi |s]] \star   \Phi^{-1}[ \xi: s_1, 0]$, 
so the source term can also be defined as  
${\rm{div}}E [\xi |s] \neq 0 $. 
Similarly, as the   monopole  can be defined as
  $G_{AB}[\xi, s] \neq 0$, and  
$({\rm{curl}} E[\xi|s])_{AB} \neq 0$. Now the monopole in the dual theory is 
characterized by  $({\rm{curl}} \tilde E[\eta|t])_{AB} \neq 0 $, and  the source in the dual theory 
is characterized by   ${\rm{div}}\tilde E [\eta |t] \neq 0$. 
We require that under the duality transformation, 
the source  in the original theory will appear as the magnetic monopole in the dual theory, so 
  ${\rm{div}}E[\xi|s]\neq 0$ should imply  
$({\rm{curl}} \tilde E [\eta|t])_{AB} \neq 0$. We also require that  the monopole in the original theory 
should appear as the source term in the dual theory, so  
 $({\rm{curl}} E[\xi|s])_{AB} \neq 0$ 
should imply   ${\rm{div}}\tilde E [\eta |t]\neq 0$. 
We can use the fact that $\eta (t)$  coincides with 
$\xi (s)$ from  $s = t_-$ to $s = t_+$, and write 
\begin{eqnarray}
&&\frac{\delta}{\delta \eta_M (t)} \star   \left(\omega^{-1} [\eta(t) ]\star   
\tilde E^A [ \eta|t] \star   \omega [\eta(t)]\right) \epsilon_{MANP}  
 \nonumber \\&=&  
 - \frac{2}{N} \epsilon^{ABCD}\frac{d\eta_B}{dt} \star  
 \int  D\xi ds \frac{\delta E_C [\xi|s] }{\delta \xi_M (s)}\star   \frac{d\xi_D}{ds} 
 \nonumber \\ &&  
\star   \left[\frac{d\xi^F}{ds}\star   \frac{d\xi_F}{ds} \right]^{-2} \delta(\xi (s) - \eta(t)) \epsilon_{MANP}.
\end{eqnarray}
Thus, we obtain 
\begin{eqnarray}
 &&\left(\omega^{-1}  [\eta(t) ]\star   ({\rm{curl}}\tilde E [ \eta|t]_{AB} \star   \omega [\eta(t)]\right)
 \nonumber \\ &=& - \frac{1}{N}  \int  D\xi ds 
\left[\frac{d\eta^C (t)}{d t} \star   \frac{d  \xi^D(s)}{ds}
-\frac{d\eta^D (t) }{d t} \star  \frac{d \xi^C (s)}{ds}\right]\epsilon_{ABCD}
\nonumber \\ && 
 \star   {\rm{div}}E[\xi |s]   
 \star   \left[\frac{d\xi^F}{ds}\star   \frac{d\xi_F}{ds} \right]^{-2}   \delta(\xi (s) - \eta(t)).
\end{eqnarray}
Thus,  if ${\rm{div}}E[\xi|s] =0$, then  $({\rm{curl}} \tilde E[\eta|t])_{AB}= 0$. 
As the duality is invertible,   we can also demonstrate that if 
${\rm{div}}\tilde E[\xi|s] =0$, then   $({\rm{curl}}  E[\eta|t])_{AB}= 0$. 
Thus, the sources in the original theory become monopoles in the dual theory, and the monopoles in the dual 
theory become sources in the original theory. It may be noted that we have analysed the sources and monopoles 
in both original and dual superloop theories. As both the supervector field and the dual supervector field 
are defined on the deformed superspace, both the original theory and the dual theory will have
$\mathcal{N} = 1/2$ supersymmetry.

\section{Conclusion}
In this paper,  we  have analysed a deformed four dimensional supersymmetric Yang-Mills theory using superloop space. 
The deformation broke half the supersymmetry of the original theory. Thus, as the original theory  had  
$\mathcal{N} =1$ supersymmetry, the theory after the deformation only has $\mathcal{N} = 1/2$ supersymmetry. 
We obtained the loop space variables for this deformed super-Yang-Mills theory in this deformed superspace. 
Thus, we obtained a  generalized of the ordinary superloop space in four dimensions. This deformed superloop 
space was used for constructing a 
duality which reduced to the ordinary loop space duality in absence of supersymmetry. Thus, for an abelian 
gauge theory without any supersymmetry, this duality reduced to Hodge duality. 
We demonstrated that under this duality the monopoles in the original theory became sources in the dual theory, 
and the sources in the original theory became monopoles in the dual theory. 

The loop space  duality for ordinary Yang-Mills theory has been used for studding various 
interesting physical phenomena \cite{1p}-\cite{d2}. 
It will be interesting to use the deformed superloop space duality constructed here, for analysing 
similar phenomena in the deformed supersymmetic theories. Thus, we can construct a deformed 
 supersymmetric Dualized Standard Model. This  deformed 
 supersymmetric Dualized Standard Model will have $\mathcal{N} = 1/2$ supersymmetry. The  
  phenomenological consequences of this model can also be studied. It will be interesting to 
analyse the ABJM theory using this deformed superloop formalism  \cite{abjm}. Furthermore, it will also
be interesting to study the effect of monopoles in the 
  ABJM theory using this formalism. It may  be noted that it is expected that the supersymmetry of the 
  ABJM will get enhanced due to monopole  \cite{enha}-\cite{enha1}. 
  It may be noted that the loop space formalism for the 
  ABJM theory has already been constructed \cite{ls}. 
  It will be interesting to study these effects 
  in the formalism developed in this paper.

\end{document}